\documentclass[jgrga]{agutex}
\usepackage{color}
\usepackage{graphicx}
\authorrunninghead{T. ZIVKOVIC, S. C. BUCHERT, H. OPGENOORTH, P. RITTER, H. L\"{U}HR}

\titlerunninghead{SMALL-SCALE FIELD-ALIGNED CURRENTS}

\begin{document}

\title{Evidence of small-scale field aligned current sheets from the
  low and middle altitude cusp continuing in the ionosphere}

\authors{T.  \v{Z}ivkovi\'{c} \altaffilmark{1},
S. C. Buchert \altaffilmark{1}, H. Opgenoorth \altaffilmark{1}, P. Ritter \altaffilmark{2} and H. L\"{u}hr \altaffilmark{2}}

\authoraddr{T.  \v{Z}ivkovi\'{c}, 
(tatjana.zivkovic@irfu.se)}

\altaffiltext{1}{Swedish Institute of Space Physics, University of Uppsala, Sweden} 
\altaffiltext{2}{Deutsches GeoForschungsZentrum, GFZ, Telegrafenberg, D-14473 Potsdam, Germany}

%
%


\begin{abstract}
  We investigate kilometer-scale field-aligned currents that
  were detected both in the magnetospheric cusp at a few Earth radii
  altitude and in the ionosphere by satellites that were, according to
  the Tsyganenko model, within a few tens of kilometers and minutes on
  the same magnetic field line. Also thermosphere up-welling that
  often accompanies the dayside field-aligned currents in the inner
  cusp was seen.  We used Cluster and CHAMP satellites, and searched
  for conjunctions during the whole year of 2008, as then the Cluster
  spacecrafts were mostly at mid-altitudes when crossing the 
  cusps.  We focus on two case studies from this period. Evidence is
  presented that sheets of small scale field-aligned current continue
  through the low altitude cusp and ionosphere. The ionospheric
  current densities are not particularly strong, a few $\mu$~Am$^{-2}$
  at about 340 km, and several tens of nAm$^{-2}$ at about 20000 km,
  implying that these currents might be relatively common events, but
  are hard to detect due to rareness of suitable locations of at least
  two satellites from different missions.
 \end{abstract}

\begin{article}
\section{Introduction}
The interaction between the near-Earth space and the ionosphere and
upper atmosphere involves field-aligned currents (FACs) as first
envisaged by \cite{B13}. The relatively large scale FAC systems
of the nightside aurora have been studied and analyzed numerously, see
for example, a review by \cite{B82}. The appearance of optical
aurora often suggests also the presence of small scale, less than a km
wide FACs. These have indeed been found with the help of sufficiently
rapidly sampling magnetometers on board of satellites in low orbits, for
example, see \cite{L94}. 
The {\O}rstedt and CHAMP satellites, launched in 1999 and 2001,
respectively, featured high-precision and fast sampling (10 and 50
Hz) fluxgate magnetometers as well as polar orbits. Perhaps a bit
surprising, the most intense FACs were found with these satellites on
the dayside in the cusps, reaching a few hundred $\mu$ A/m$^{2}$ and
with scales down to a few hundred meters \citep{NC2003, W2003}.
Particularly, it was shown by \cite{NC2003}, who used {\O}rsted, that
small-scale FACs were 1--2 order of magnitude larger than large scale
region 1 and 2 currents. 
CHAMP is a German mission, which was
flying on the altitude between 300-450 km with an inclination of
$87.3^{\circ}$. Besides the Earth's magnetic field, CHAMP, was designed to map the Earth's gravity
field and carried also an accelerometer, and was accurately tracked with
GPS and satellite laser ranging. This allowed to determine accurately
the satellite's air drag.
\cite{L2004} showed that air drag plots
from CHAMP had one dominant oscillation which was due to the air
density difference on the day and night side of the
Earth. Superimposed on this oscillation were relatively narrow peaks in air drag
corresponding to widths of a few ten to hundred kilometers of enhanced neutral density.
These thermospheric upwellings seemed to occur in the cusps and were
accompanied by intense kilometer scale FACs, which were obviously causing the
upwellings.

\noindent

There are at least two obvious questions that deserve attention: what
is the origin of the intense FACs in the cusps, and how can they cause
thermospheric upwelling. In this paper mainly the first question is
addressed. 
\cite{W2008} used, in addition to magnetometer measurements of {\O}rsted
and CHAMP, particle data from the low orbiting DMSP satellites. He
showed that cusp locations as inferred from magnetosheath-like
particle precipitation matched well the locations of small-scale
currents, but FACs seem to be generated also in the transition zone
between the low-latitude boundary layer (LLBL) and the cusp. Small
scale intense FACs could also, with help of the DMSP F15 ion drift and
magnetometer, be associated with very strong enhancements of the
Poynting flux \citep{L2011}. These authors proposed that magnetic
reconnection in the cusp, particularly during a large $B_Y$ component
and northward $B_Z$, is a cause of such small scale, but intense Poynting
flux increases.
 \cite{R2007} proposed an explanation of the
FACs in terms of Alfv{\'{e}}n waves trapped in a ionospheric
resonator. Then one would expect that the FACs which are observed
localized in the ionosphere cannot be mapped to regions above the
trapping resonator.  Also, at higher altitudes, several Earth radii,
localized strong FACs have been found in satellite data
\citep{J2010}. However, observations with direct evidence
for ionospheric signatures of high altitude small scale FACs are rare.
Here we use the Cluster spacecrafts and the CHAMP satellite to
investigate whether the continuation of ionospheric small-scale
current sheets can be found at much higher altitudes of a few Earth
radii. The four ESA Cluster satellites were launched in 2001 with an
initial perigee at 26000 km (4 RE) and apogee at 124000 km (19 RE)
with orbits that relatively frequently crossed the cusps at varying
altitudes.
Previously, in so far only one event intense FACs were seen nearly
simultaneously both in the ionosphere as well as at about 10 Earth
radii distance in the outer cusp \citep{B2012}. The mapping of the
magnetic field using the Tsyganenko model \citep{T96} suggested, that
the FAC seen by CHAMP might have continued all the way, and the
Cluster measurements further suggested that the cause of the FAC is
related to nearby ongoing magnetic reconnection.  In this paper, we
are particularly interested in cases where the Cluster is located 
at mid altitudes (a few $R_E$), and where chances to
find FAC sheets extending over large distances from the ionosphere
into near Earth space are greater than when Cluster is located in the
exterior cusp.

\section{Results}
We have searched for conjunctions between Cluster C3 and CHAMP
spacecrafts. The following criteria for a conjunction were adopted:
within a time period of at most 20 minutes the difference between the
geographic latitudes of the mapped Cluster footpoint and CHAMP is at
most 0.5 degrees, while the difference between the geographic
longitudes at most 20 degrees. Of particular interest were the
conjunctions that occurred at mid altitudes, when a Cluster satellite
was at about 2-3 $R_{e}$. In 2008, Cluster spacecrafts
were far apart from each other and it was not possible to use results
from more than one Cluster spacecraft but we have selected this year
since Cluster spacecrafts crossed frequently the mid-altitude cusps.

In order to compute Cluster footprints we have used the Orbit
Visualization Tool (OVT), which called the Tsyganenko 96 model
\citep{T96} for the field line tracing.  OVT is described and can be
downloaded from http://ovt.irfu.se. 
The Cluster footprints are computed for an altitude of 120 km above the
Earth, with 1 min time resolution. The CHAMP spacecraft's circular
orbit was at about 340 km altitude, and its orbital velocity
was $v_{s}=7.7$~km s$^{-1}$.  We have detected 11 conjunctions between
Cluster C3 and CHAMP at mid altitudes in 2008, but some of these
conjunctions revealed relatively weak FAC on the CHAMP satellite,
probably because no intense FAC sheets were present at this time, or
the CHAMP satellite orbit missed then.
To investigate the interesting conjunctions further, the
CHAMP fluxgate magnetometer data were used to estimate current
densities. Neutral density estimates were obtained from the TU Delft thermosphere web server
(http://thermosphere.tudelft.nl/acceldrag/index.php). The processing
is described in \cite{D2010}. Further, we use magnetic field data from the Cluster FGM
instrument \citep{B2001}, electric field data from the EFW instrument
\citep{G2001}, spin resolution electron spectrograms from the Plasma
Electron and Current Experiment (PEACE) \citep{J97}, as well as ion
spectrograms of the precipitating particles on the Cluster CIS
instrument \citep{r2001}. We were then left with two conjunctions for
the discussion.

\begin{figure}
\begin{center}
\includegraphics[width=7cm]{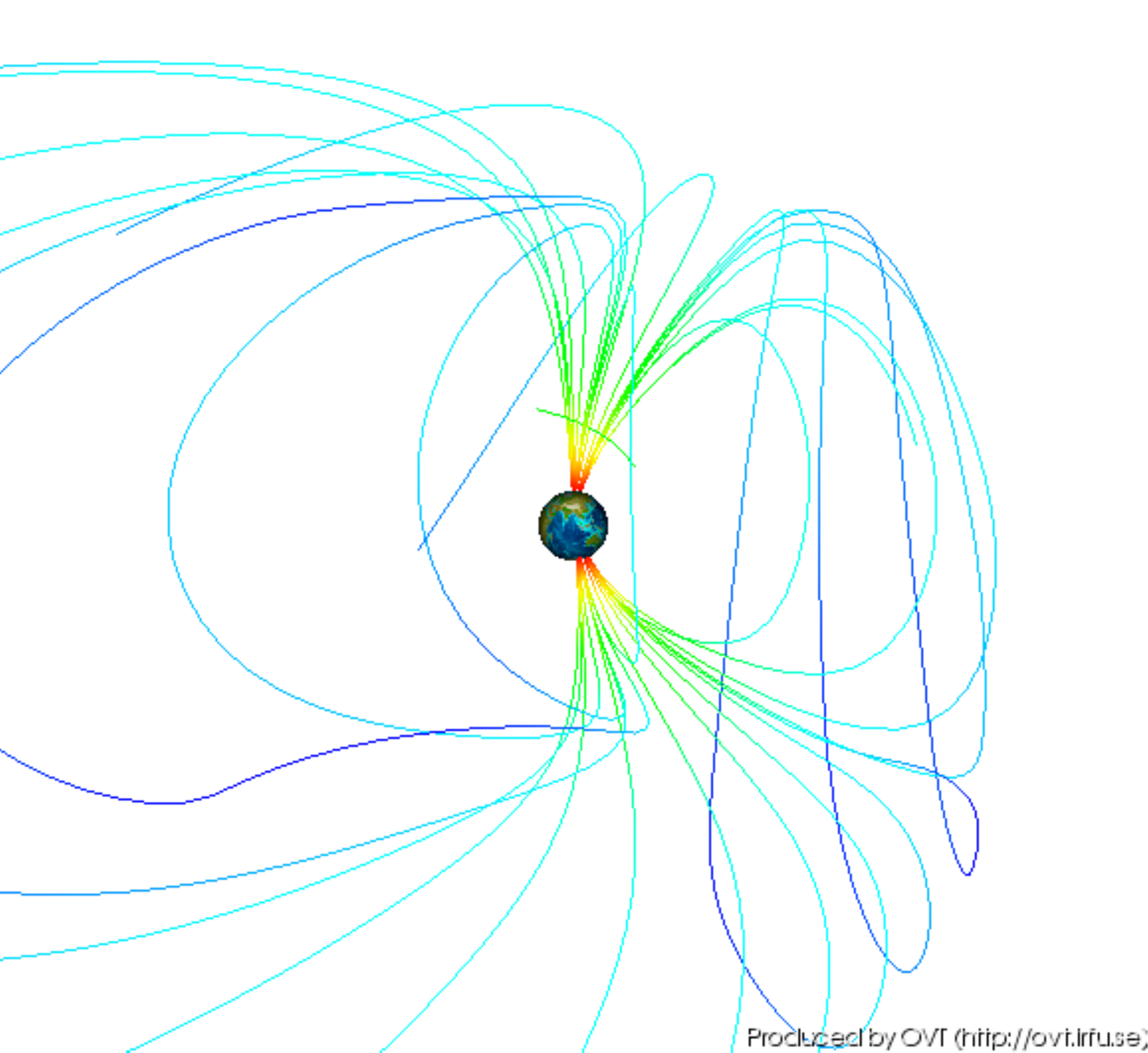}
\caption{The 3-dimensional magnetic field according to Tsyganenko 96
  during the conjunction on July 17, 2008 is plotted, the orbit of the
  Cluster C3 satellite is shown by a green line.}
\end{center}
\end{figure}

\begin{figure}
\begin{center}
\includegraphics[width=7cm]{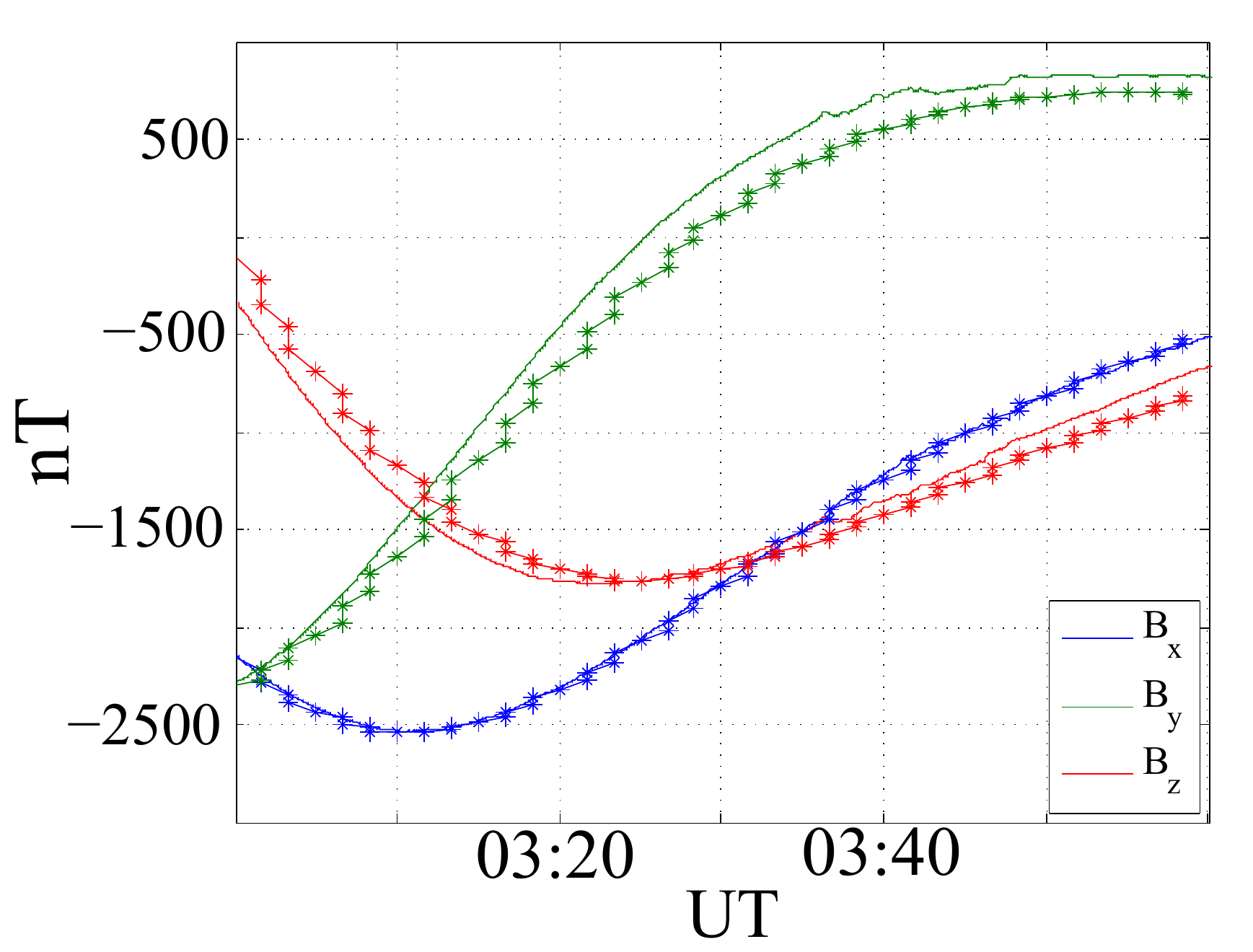}
\caption{Magnetic field in GSE coordinates from Tsyganenko model (stars) and Cluster C3 FGM instrument.}
\end{center}
\end{figure}

\begin{figure}
\begin{center}
\includegraphics[width=6cm]{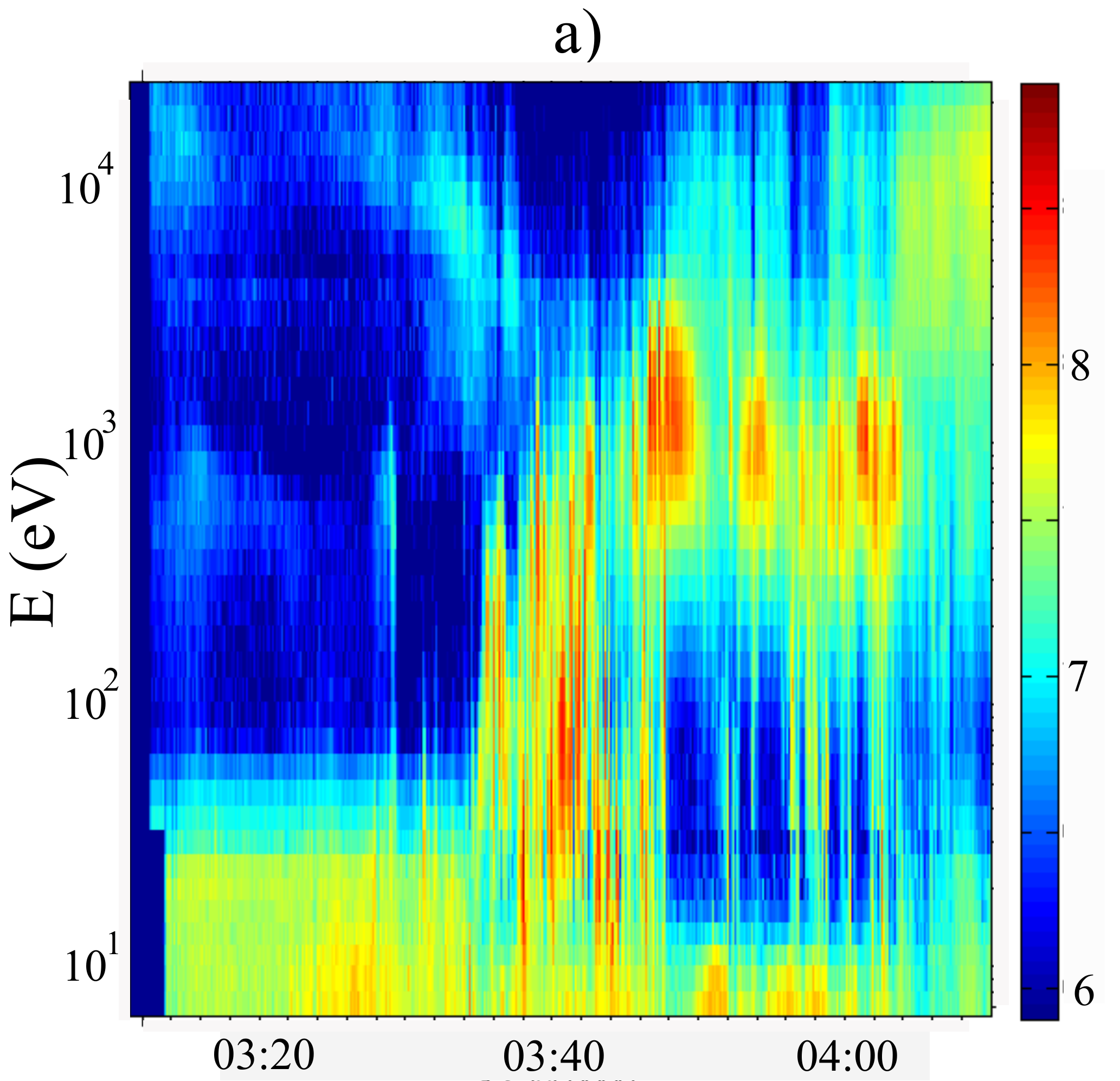}
\vspace{0.2cm}
\includegraphics[width=6cm]{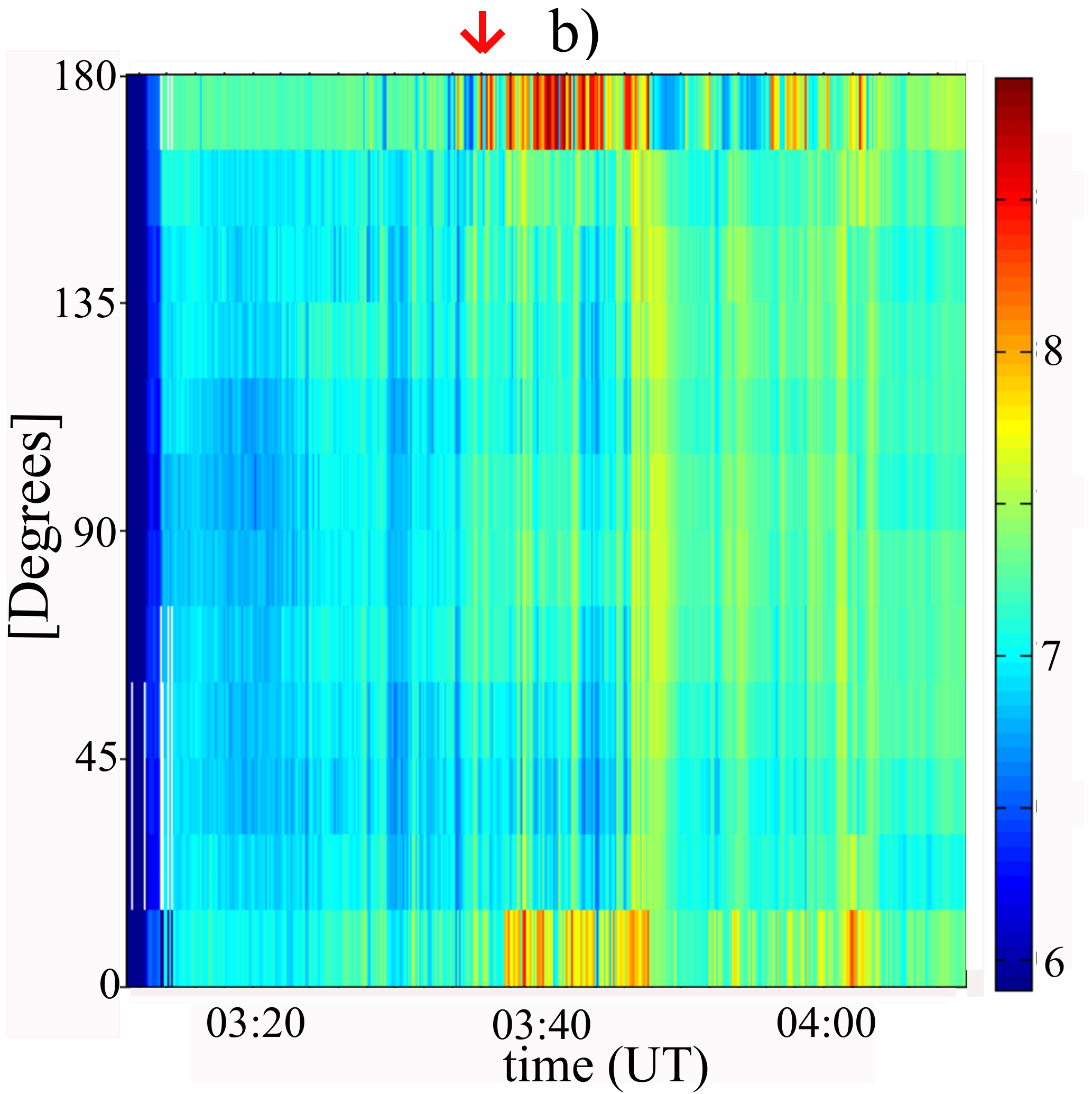}
\caption{Electron data from the PEACE instrument in C3: a) Electron
  energy flux, b) Electron angular distribution (red arrow indicates current density from Figure 6.}
\end{center}
\end{figure}

\begin{figure}
\begin{center}
\includegraphics[width=6cm]{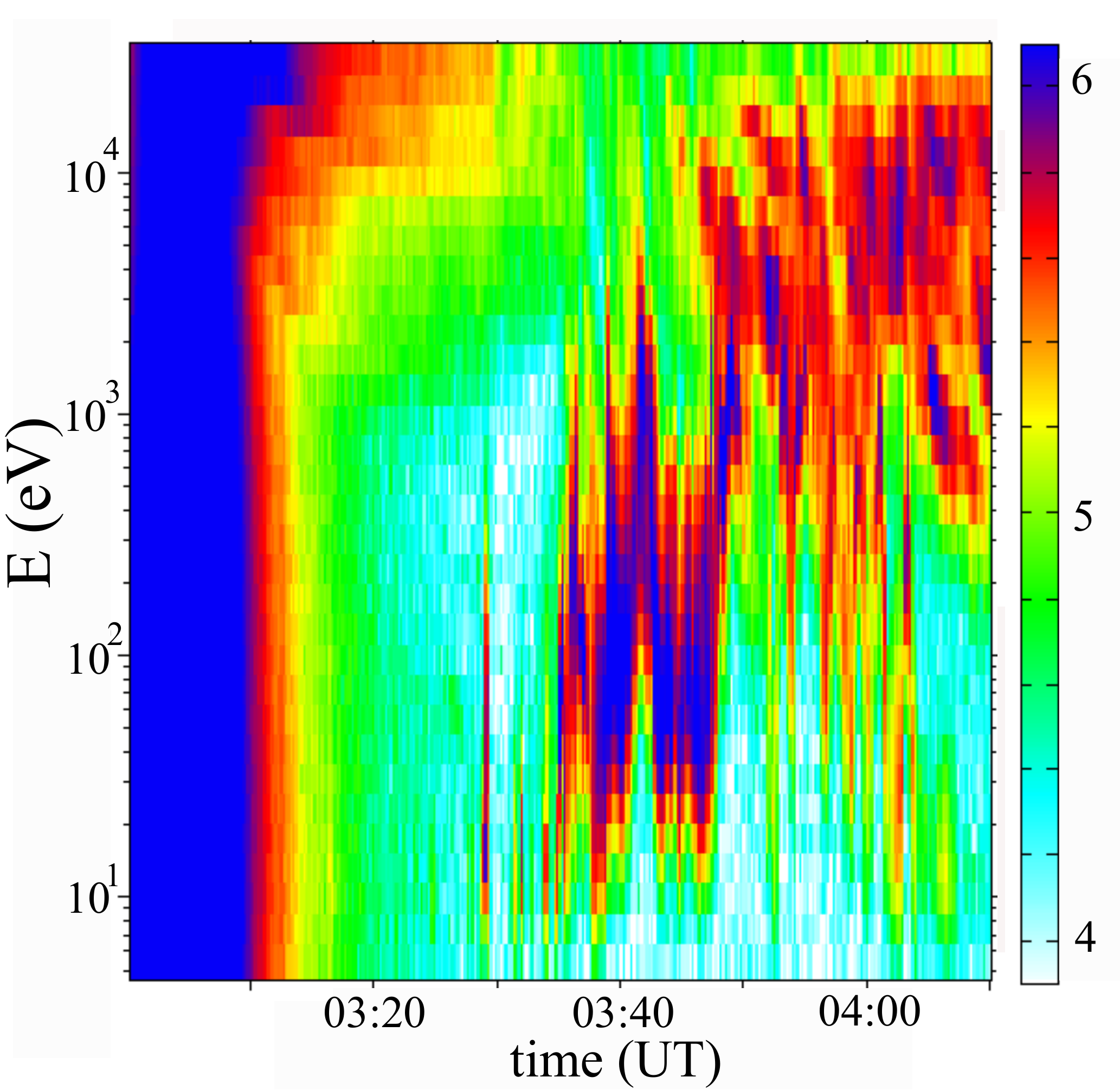}
\caption{Ion energy flux according to the CIS-HIA instrument.}
\end{center}
\end{figure}

\begin{figure}
\begin{center}
\includegraphics[width=8cm]{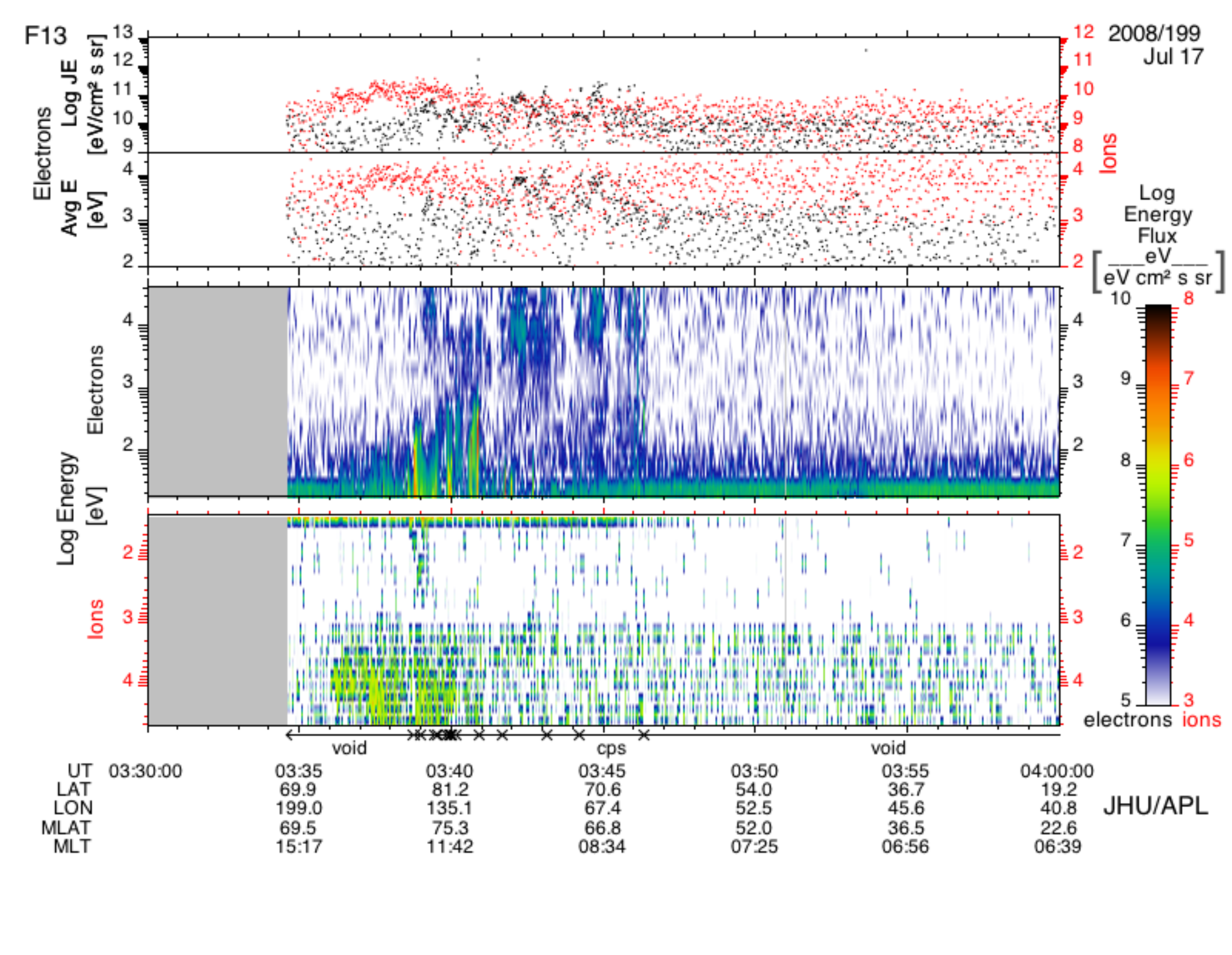}
\caption{Ion and electron spectrograms from DMSP F13 on July 17, 2008.}
\end{center}
\end{figure}

\begin{figure}
\begin{center}

\includegraphics[width=7cm]{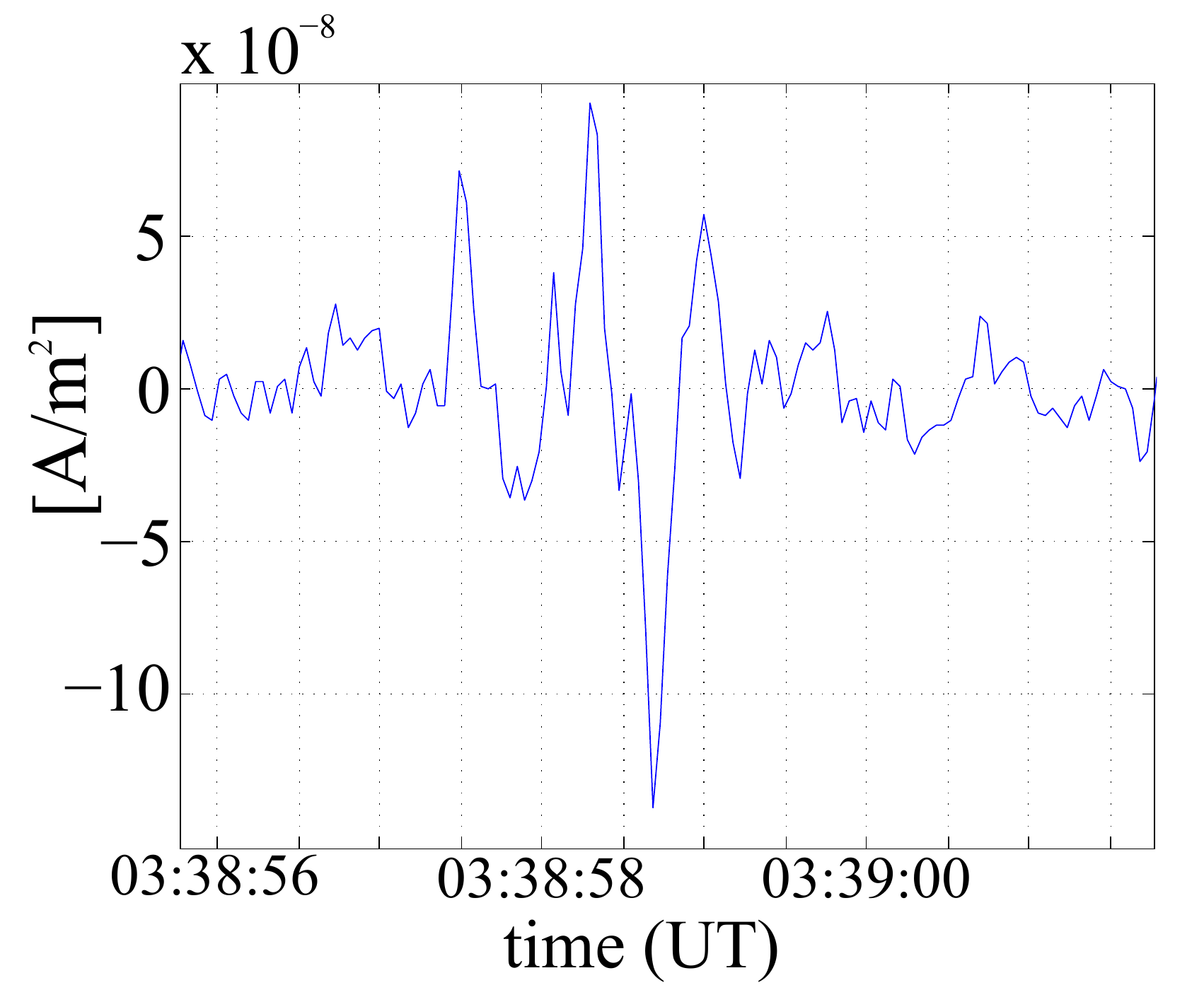}
\caption{Current density estimates for Cluster C3.}
\end{center}
\end{figure}

\begin{figure}
\begin{center}
\includegraphics[width=6cm]{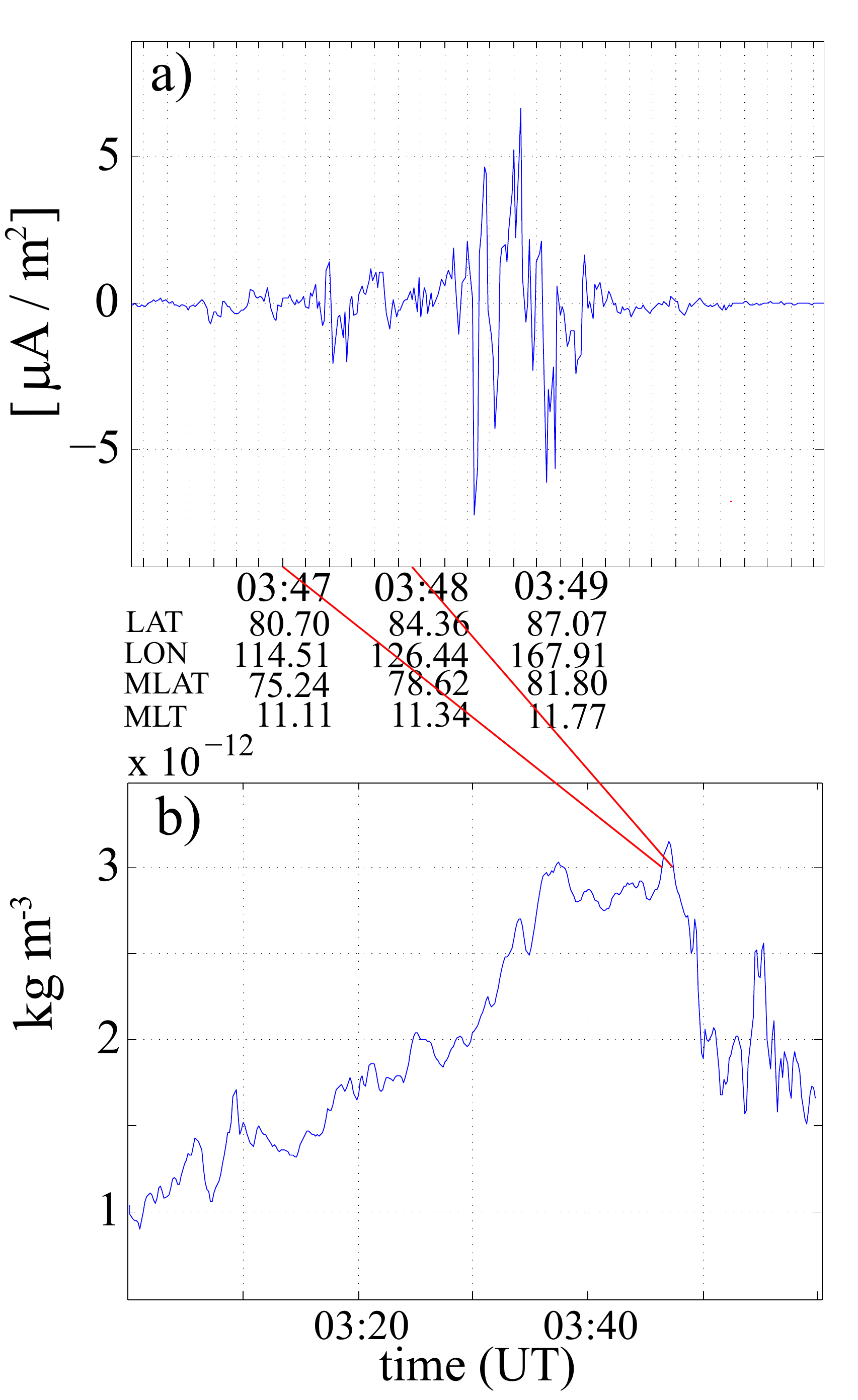}
\caption{Estimates of a) Current density, and b) Neutral
  density, from the CHAMP magnetometers and accelerometers,
  respectively.}
\end{center}
\end{figure}

\begin{figure}
\begin{center}
\includegraphics[width=7cm]{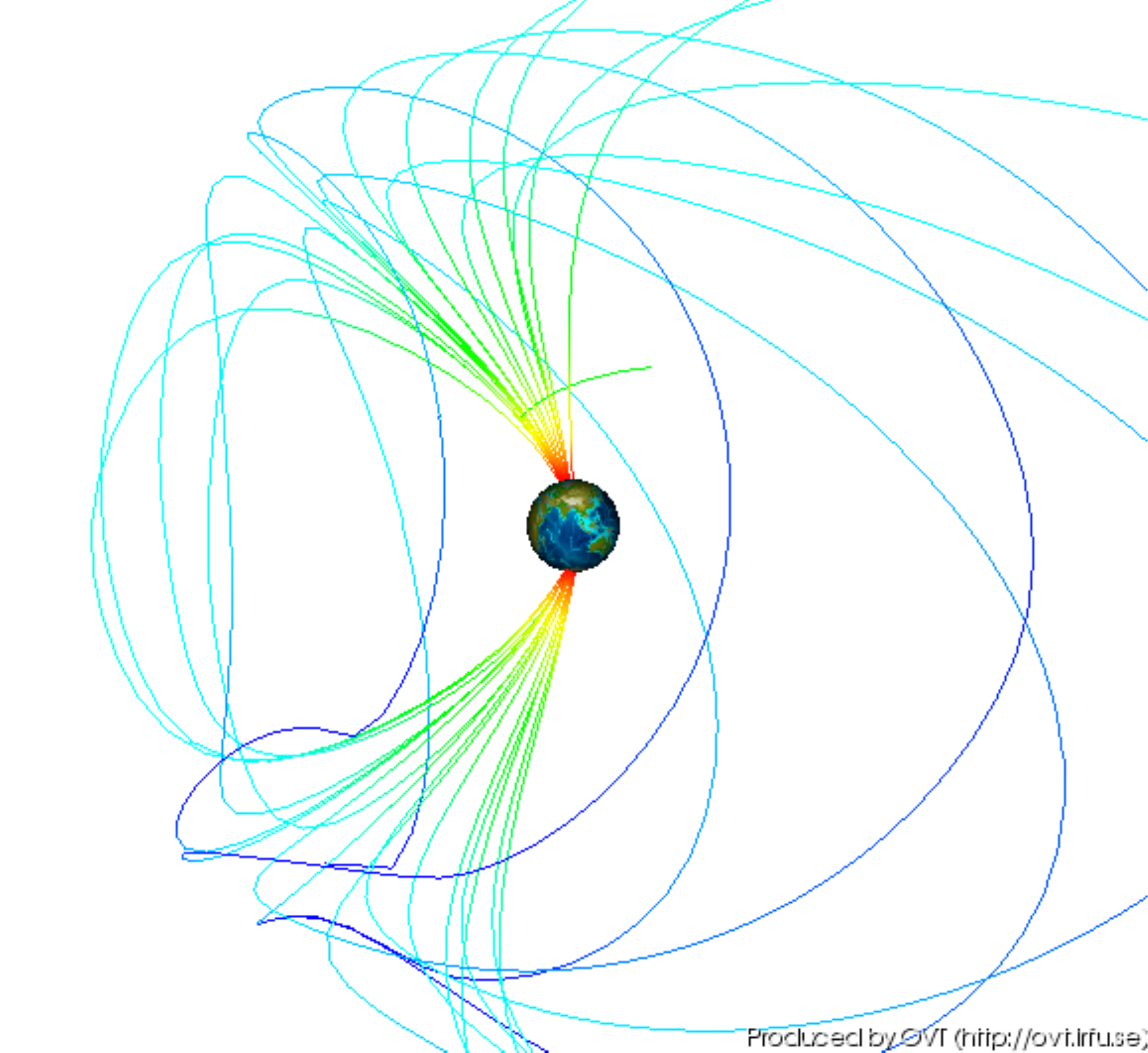}
\caption{The 3-dimensional magnetic field according to Tsyganenko 96
  during the conjunction on August 7, 2008 is plotted, the orbit of the
  Cluster C3 satellite is shown by a green line.}
\end{center}
\end{figure}

\begin{figure}
\begin{center}
\includegraphics[width=6cm]{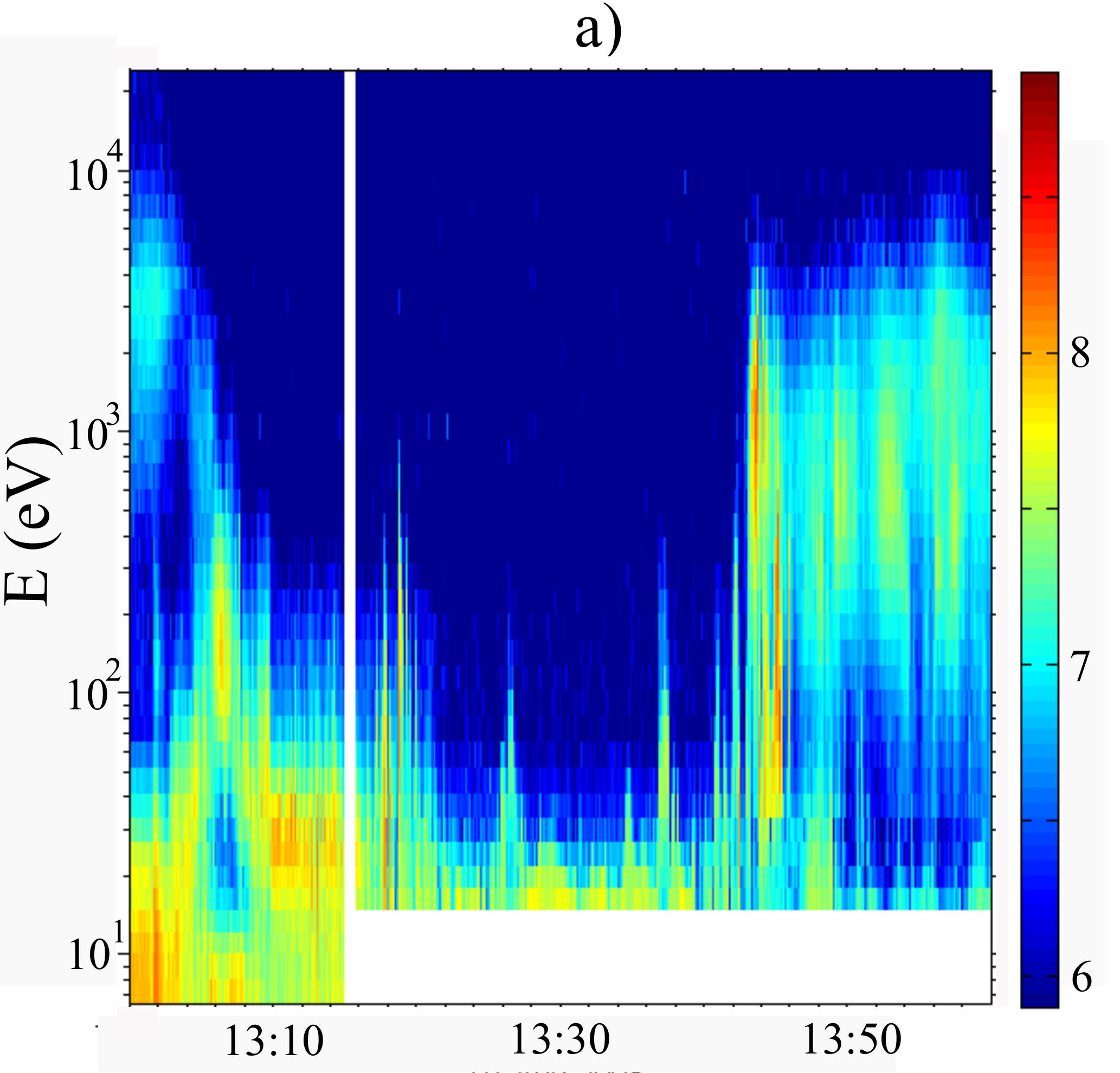}
\vspace{0.2cm}
\includegraphics[width=5.7cm]{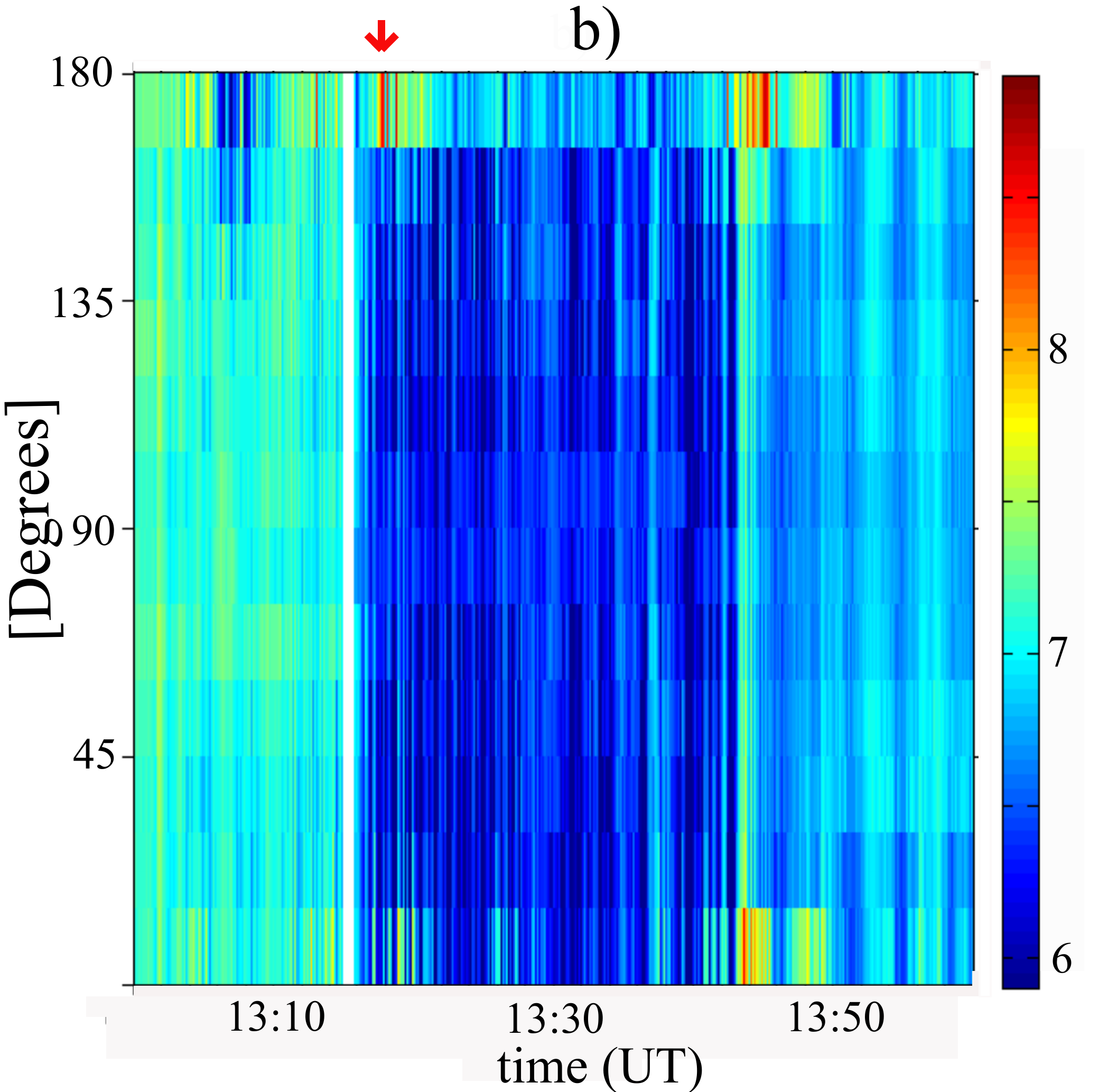}
\caption{Electron data from the PEACE instrument in C3: a) Electron
  energy flux, b) Electron angular distribution (red arrow indicates current density from Figure 11).}
\end{center}
\end{figure}

\begin{figure}
\begin{center}
\includegraphics[width=6cm]{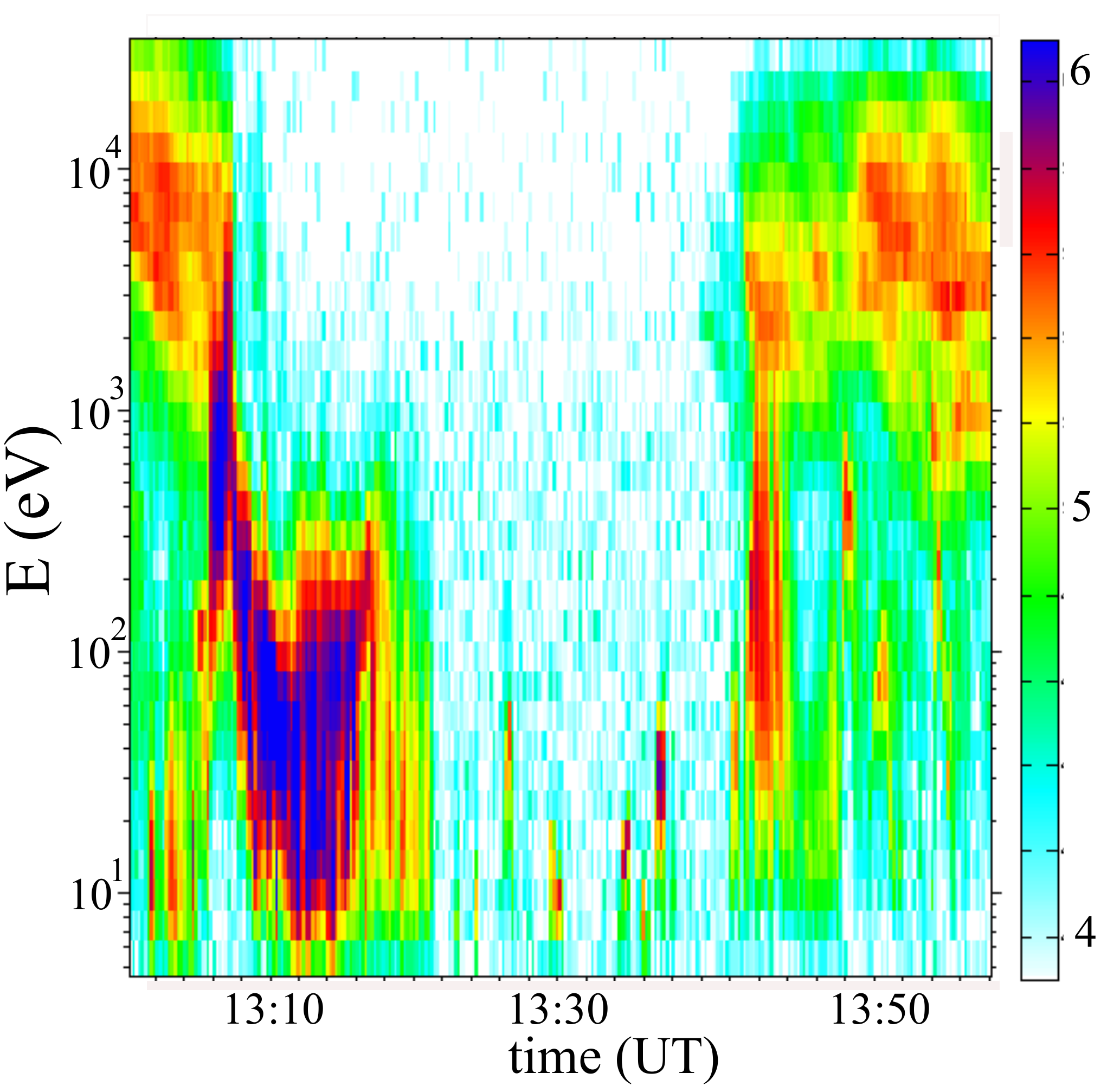}
\caption{Ion energy flux according to the CIS-HIA instrument.}
\end{center}
\end{figure}

\begin{figure}
\begin{center}

\includegraphics[width=6cm]{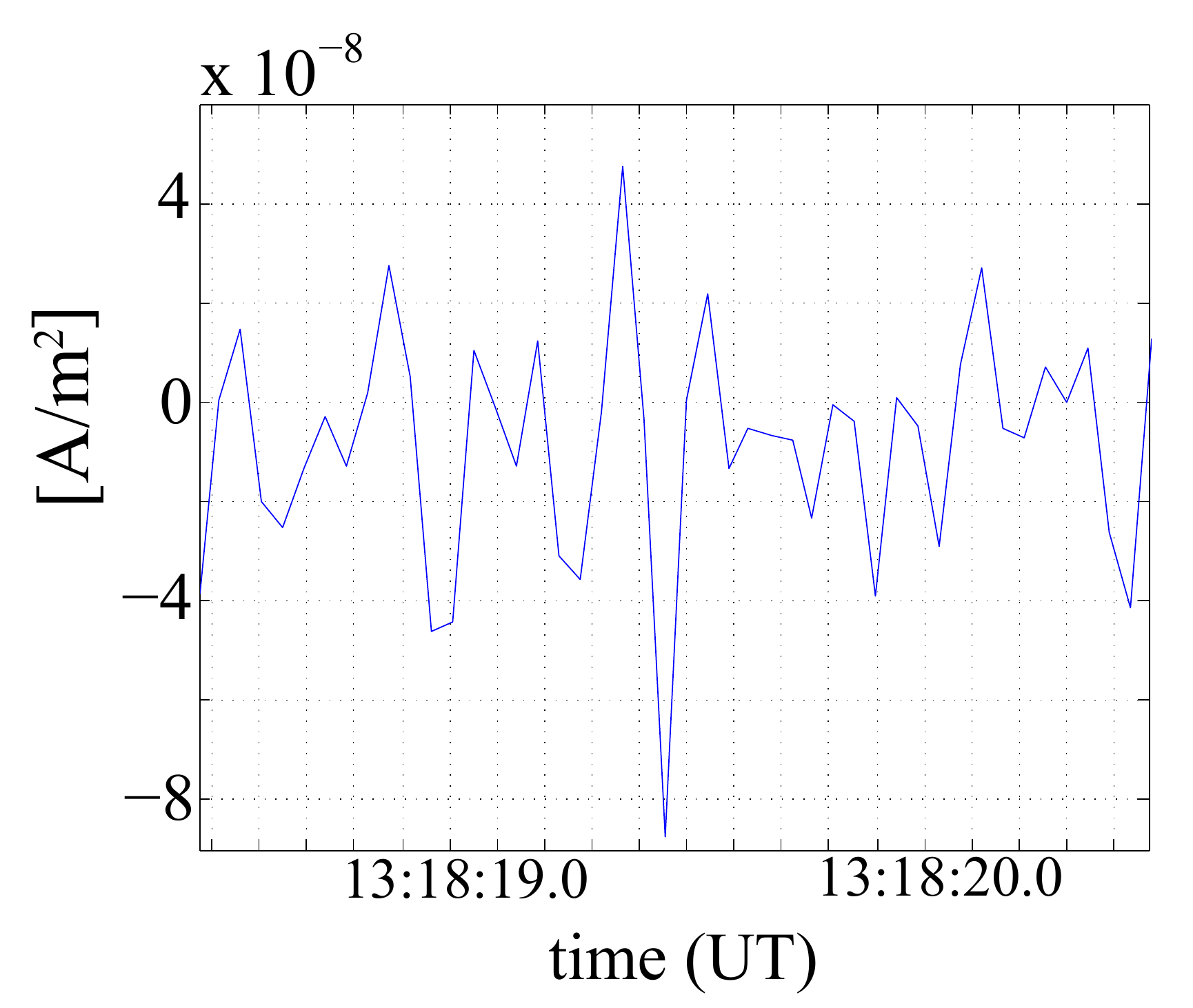}
\caption{Current density estimates for Cluster C3.}
\end{center}
\end{figure}

\begin{figure}
\begin{center}

\includegraphics[width=6cm]{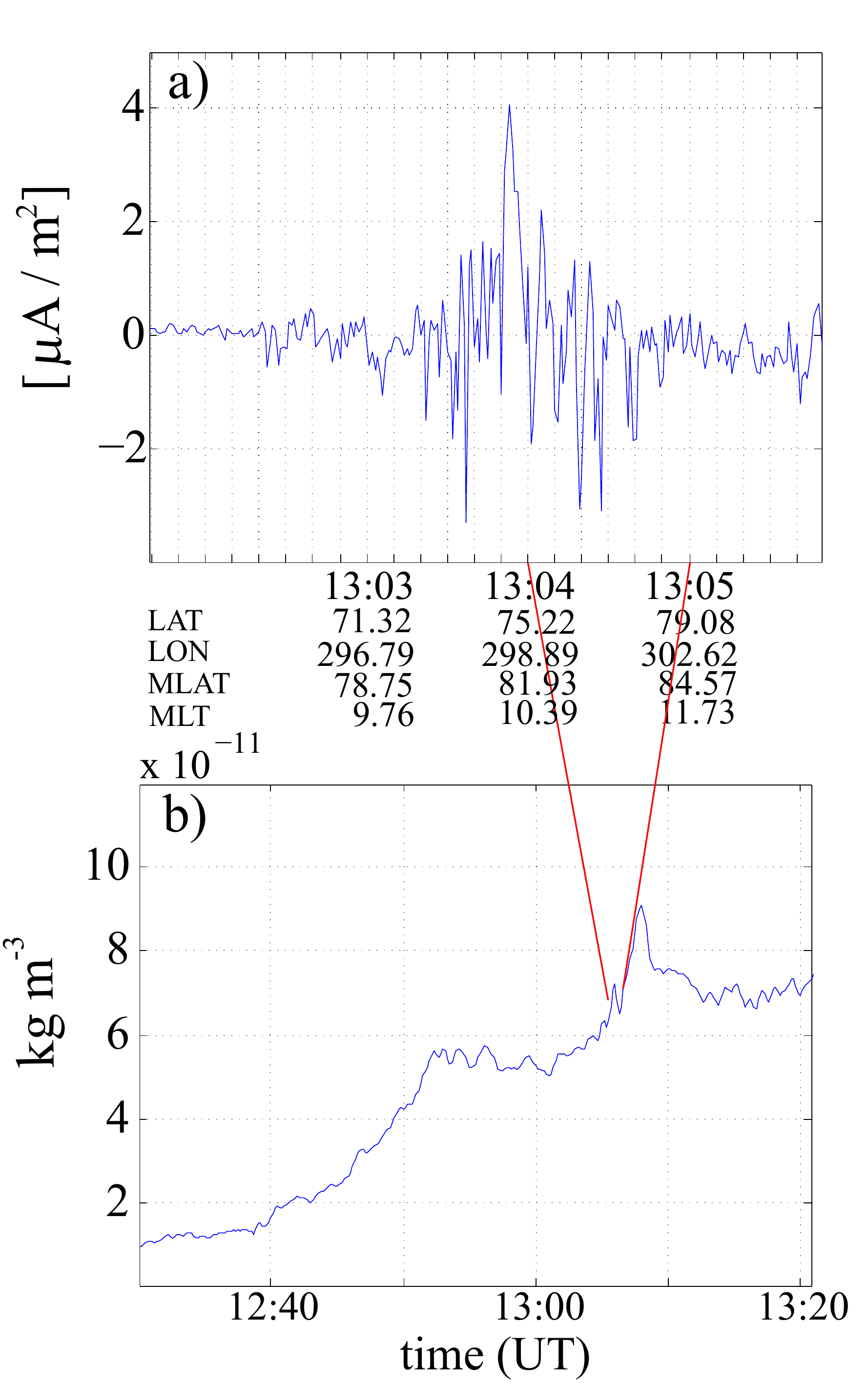}
\caption{Estimates of a) Current density, and b) Neutral
  density from the CHAMP magnetometers and accelerometers,
  respectively.}
\end{center}
\end{figure}

\subsection{July 17, 2008 Cusp event}
During this event the Cluster C3 spacecraft entered, according to the
ion particle data (confirm Figure 4), the cusp of the Northern
hemisphere at about 03:30 UT. The Tsyganenko model predicted the
C3 would be on open field lines from 03:20 to 03:40 UT. Its
passage is shown as a green line in Figure 1.
%
C3 was then at an altitude of 2.8 $R_e$, and the mapped geographic
coordinates at 03:30 UT, were 81.5 north latitude and 135.79 east longitude (MLAT: 75.24 N, MLT: 11.65). The
closest other Cluster spacecraft, C1, arrived to the same area of the
inner cusp 30 min earlier and, hence, could not be used for our study.
At 03:47 UT, seventeen minutes after the Cluster C3 spacecraft, the
CHAMP spacecraft had arrived, at 81.82 north latitude and 116.77
east longitude (MLAT: 75.24 N, MLT: 11.10).  The OMNI database showed that the interplanetary
magnetic field (IMF) for the period of conjunction was relatively
quiet: $B_{x}$ was varying between -1 and 1~nT; $B_{y}$ was negative
all the time varying between -2.2~nT and -0.3~nT, while $B_{z}$ was
slightly negative, varying around $\sim -1.5$ nT. It became positive
after 03:25~UT reaching maximum value of 0.7 nT. In addition, the
Solar wind flow pressure was at about $\sim 1.3$ nT, and the
$D_{st}\sim -13$.  In order to get as precise footprints as possible,
we have used the IMF, flow pressure and $D_{st}$ index starting two
hours before the event, as the input parameters to the Tsyganenko
model. Figure 2 shows the magnetic field at Cluster C3 and the one
predicted from the Tsyganenko model at the altitude of Cluster, both
in GSE coordinates. Apart from a slight disagreement in the $B_{Y}$
component, the Tsyganenko model gives good description of the magnetic
field at Cluster.
Figure 3a shows for the period of 03:00 to 04:10 UT the energy flux of
electrons estimated from the PEACE instrument, and in Figure 3b the
pitch angle distribution of the electrons is displayed.  We see energies up to
$1$keV with fluxes up to $10^{9}$ eV cm$^{-2}$ s$^{-1}$ sr$^{-1}$
eV$^{-1}$, which is typical for the cusp. In the period between 03:38
and 03:48 UT
electrons are mostly
traveling down to the ionosphere (at $180^{\circ}$), though a
significant
population of electrons is also traveling out of the ionosphere (at
$0^{\circ}$) along the magnetic field lines.
In Figure 4 ion spectrograms from Cluster C3 for the time between
03:00:00 and 04:10:00 UT are shown.  High fluxes of $\sim$ keV ions
are typical for the polar cusp, which gets filled by magnetosheath
ions.  In the same figure, a reversed ion dispersion can be seen,
which is consistent with the IMF measurements. Namely, since IMF
$B_{z}$ is positive, while IMF $B_{x}$ and $B_{y}$ are small, a
reconnection might have occurred in the lobes, poleward of the cusp
\citep{P2012}. The convection of the magnetic field lines is then
sunward and ion energy decreases with decreasing latitude which causes
a reverse ion dispersion.  This is consistent with ion velocity
measurements on HIA instrument on CIS, where positive $V_{x}$ confirms
that magnetic field lines convect sunward.  Similar fluxes and
energies of electrons and ions are measured in the same time interval
on the DMSP satellite F~13, whose altitude at the time was about 850
km, and geographic coordinates were 81.2$^\circ$ latitude and
135.1$^\circ$ longitude. The DMSP particle spectrograms are shown
in Figure 5.

From the FGM instrument on Cluster C3, we have estimated the parallel
current density from 03:38:56 to 03:40:00~UT. In Figure 6, we see a
positive, upward current density peak of about $90$ nAm$^{-2}$ and
then negative peak of about $140$ nAm$^{-2}$. During
this particular cusp crossing these current density peaks were the
largest ones. The footprints for this current are 83.03$^\circ$ north
latitude and 100.43$^\circ$ east longitude (MLAT: 76.20 N, MLT: 10.38).
The signs of the current density peaks, first positive, then negative
agree with the electron particle data from PEACE, see Figure 3b, where
electrons first stream down to the ionosphere (an upward current) and
then out of the ionosphere (a downward current).  The current
was estimated from the velocity of the current sheet relative to the
C3 spacecraft and the magnetic field on the Cluster 
using Ampere's law, $\nabla \times {\bf B}=\mu_0 {\bf j}$ in a finite
difference approximation: $\mu_0 {\bf j} = {\bf v}/(|{\bf v}|^2dt)
\times {\bf dB}$. The FAC density is then ${\bf j}\cdot {\bf B}/|{\bf
  B}|$. Here ${\bf dB}$ is the difference of the magnetic field vector
over the sampling interval $dt$, which is about $1/(22 \mathrm{Hz})$.
The velocity components of ${\bf v}$ are those measured by the ion CIS
instrument. We assume that ions are moving mostly transverse to the
field-line direction, and the current sheet is frozen into the
plasma. Assuming that the current sheet was stationary while Cluster
was passing, a current sheet width of 43~km could be estimated.  The
level~2 fluxgate magnetometer data from CHAMP have a time resolution
of 1~s with a nominal amplitude resolution of 0.1~nT. The FAC density
in the ionosphere is estimated along the satellite track from the
magnetic field component perpendicular to the auroral oval, assuming
stationary sheet currents \citep{WL2005}.  Positive currents flow
upward.  The estimated FAC density is shown in Figure 7a. Two periods
of bursts of FACs can be identified: first, around 03:47:20~UT, near
81.95$^\circ$ north latitude and 117.26$^\circ$ east longitude (MLAT: 76.37 N, MLT: 11.17), both
positive and negative currents of about 2~$\mu$Am$^{-2}$ occur. A
second burst happens at 03:47:45~UT near 83.48$^\circ$
north latitude and 122.20$^\circ$ east longitude (MLAT: 77.78 N, MLT: 11.27). The latitude and
longitude of the mapped current seen by Cluster C3 at 03:38:58~UT is
close to the coordinates of CHAMP at the negative peak in this second
burst.  However, we believe that really the first burst at CHAMP
corresponds to the current seen by Cluster C3. Firstly, the ion and
electron fluxes at DMSP F13 (see Figure~5), which agree reasonably
with the mapped fluxes at Cluster, are seen at a latitude close to the
first current increase at CHAMP. Secondly, the small but
significant difference between the Tsyganenko modeled and by Cluster measured magnetic field
at 03:38~UT (see Figure~2) suggests that the footprint of Cluster is
more southward than obtained from the Tsyganenko field-line
tracing. Finally, we see both a positive and a negative current on
Cluster and that agrees better with the positive and negative currents
at CHAMP seen in first burst of FAC. Further, we can estimate how well
the intensity of the current density on CHAMP corresponds to the
measured current density at the Cluster C3 spacecraft. Assuming that
FAC does not close between the ionosphere and Cluster altitude, and
also does not connect to other transverse currents like for example
polarization currents in Alfv{\'{e}}n waves, the current density in
the ionosphere $j_{i}$ should be approximately $B_{i}/B_{cl} j_{cl}$,
where $B_{i}$ is the ionospheric magnetic field, $B_{cl}$ is the mean
magnetic field on Cluster and $j_{cl}$ is the estimated current
density at Cluster. In our case, the current density on Cluster has a
positive spike of $j_{cl}\approx 9\cdot10^{-8}$~Am$^{-2}$, which is
followed by a more pronounced, negative spike of $j_{cl}\approx
-1.4\cdot10^{-7}$~Am$^{-2}$.  For $B_{i}\approx 50 \mu$T and a mean
magnetic field on Cluster $B_{cl}$ of about 2 $\mu$T, the estimated
current density in the ionosphere, should be for the positive spike
$j_{i}\sim 2\cdot 10^{-6}$~Am$^{-2}$ and for the negative spike
$j_{i}\sim -3.35\cdot10^{-6}$~A/m$^{-2}$, which is slightly higher
than values measured by CHAMP. If the current structure originates
in the magnetosphere (for example from magnetic reconnection) then
transverse diffusion of current by ion-electron collisions or
wave-electron interaction would cause a widening of the structure in
the ionosphere and a decrease in the current density at CHAMP.
Due to the magnetic field convergence, the width of the relevant
current sheet, $l$, which is 43~km at Cluster altitude, should get
concentrated in the ionosphere to $l\cdot \sqrt{2 \mu\mathrm{T}/50
  \mu\mathrm{T}}\approx $~8.6 km. Next, we plot normalized neutral
density, which is obtained from the recently calibrated accelerometer
dataset from CHAMP \citep{D2010}.  The density is projected to 325 km
altitude as $\rho(325 km)=\rho(h)\cdot \rho_{m}(325
km)/\rho_{m}(h)$. Here, $\rho_{m}(h)$ is the model density according
to the NRLMSISE-00 atmospheric model \citep{p2002} at the satellite
altitude, while $\rho$ is the measured density at the same altitude.
Similar to the events reported by \cite{L2004} we also see an increase
of the neutral density in the time around 03:47 UT, as shown in Figure
7b.

\subsection{August 7, 2008 Cusp event}
During this event the Cluster C3 spacecraft, according to the ion
spectrometer (compare with Figure 10), exited from the cusp of the
Northern hemisphere at about 13:20~UT (and perhaps reentered later at
about 13:30~UT). The Tsyganenko model predicted that C3 would be on
open field lines from 12:50 to 13:25~UT. C3 was at the altitude of
about 2.60 $R_{e}$. The orbit is indicated with a green line in
Figure 8.  Here, the CHAMP spacecraft arrived first, at 13:05~UT,
within the area where according to our criteria a conjunction
occurred. The geographic coordinates were latitude 79.08$^\circ$ and
longitude 302.62$^\circ$ (MLAT: 84.57 N, MLT: 11.73).
The best conjunction of Cluster~3 was at 13:18 UT, 79.16$^\circ$
latitude and 292.44$^\circ$ longitude (MLAT: 86.60 N, MLT: 11.08). The closest passage of any
other Cluster satellite, C1, occurred 40 min earlier, and hence data
from this spacecraft could not be used in our study. The IMF data from
the OMNI database showed that $B_{x}\approx -2$nT, $B_{y}\approx -4$nT
and $B_{z}\approx -1$nT, flow pressure $\approx$ 0.8, and
$D_{st}\approx 1$.  The electron spectrogram in Figure 9 a) shows
energies of up to 800 eV, with fluxes of up to $10^{9}$ eV cm$^{-2}$
s$^{-1}$ sr$^{-1}$ eV$^{-1}$, which are typical values for the
cusp. Figure 9 b) shows the electron distribution as a function of
pitch angle. One can see that the largest population of electrons
moves along magnetic field lines at around 13:15-13:20 UT, pitch angle
is $180^{\circ}$, which means that electrons go down to the
ionosphere. A smaller population of electrons has a pitch angle of
$0^{\circ}$.  From the ion spectrogram in Figure 10, we see the ion
dispersion characteristic for the southward IMF, where less energetic ions
stream along more poleward magnetic field lines when the reconnection
occurs at the dayside magnetopause. Further, we estimate the parallel
current density at C3 the same way as for the first event (see Figure
11). The strongest current density increase occurs at around
13:18:19~UT, and that is also the strongest current density increase
in the period that Cluster C3 stayed in the cusp. A burst of a few
$\mu$Am$^{-2}$ strong currents started at CHAMP (see Figure 12 a) with
the most intense current around 13:04:00~UT, at about 75.22$^\circ$
north latitude and 298.89$^\circ$ east longitude (MLAT: 81.93, MLT: 10.39). In this event, the
magnetic field of the Tsyganenko model and the one measured by Cluster
C3 deviate considerably, particularly in the $B_{Y}$ and $B_{Z}$
components (not shown here). There is no obvious explanation why the
Tsyganenko model is relatively unrealistic in this case, as the event
under study occurred under very quiet geomagnetic and solar wind
conditions. Nevertheless, moderately strong FACs are seen on both
Cluster and CHAMP within about 14~minutes in areas that are within
some uncertainty conjugate to each other.  For the ionospheric
magnetic field of $B_{i}\sim 50 \mu$T, an average magnetic field on
Cluster of $B_{cl}\sim 2.1 \mu$T and the positive current density
estimated for Cluster of $j_{cl}\sim 4.5\cdot 10^{-8}$~Am$^{-2}$, the
ionospheric current density should be $j_{i}\sim 1.13\cdot 10^{-6}$
A/m$^{-2}$. The negative current density estimated on Cluster is $j_{cl}\sim -8\cdot 10^{-8}$~Am$^{-2}$,
which gives the ionospheric current density of $j_{i}\sim -2\cdot 10^{-6}$, which agrees well with the positive and
negative current measured at CHAMP at 13:04:00~UT.
The width of current sheet at Cluster is estimated to be about
$6.5$~km. Then a sheet width of only about $1.3$~km is expected in the
ionosphere at CHAMP altitude which is below the resolution of the
available magnetic data. Nevertheless, the neutral density (normalized to 325~km)
increases also here when CHAMP crossed the region where thin current sheet
in the ionosphere was located, i.e. at around 13:05~UT. This is shown in
Figure 12b). Note that here the peak in density should be related to the FAC burst,
but one cannot expect a 1-1 correspondence because of a field-line inclination and
a motion of FAC during upwelling time.

\section{Discussion and Conclusion}
We have shown two cases of small-scale field-aligned currents which
stretch from the magnetosphere through the ionosphere inside the polar
cusp.  These current sheets are identified in the magnetosphere with
the Cluster C3 spacecraft, and in the ionosphere with the CHAMP
spacecraft. These currents seem quite "geo-effective", since they heat
the ionosphere and upper atmosphere and cause upwelling of the
thermosphere. It has been pointed out already by \cite{L2004}, that
the density enhancements accompanied by small-scale field-aligned
currents occur quite independently of magnetic activity.  Also these
events occured under only moderate or quiet geomagnetic and solar wind
conditions. A similar study has been carried by \cite{B2012}, where
the ``same'' current sheet was suggested to be seen both by Cluster
and CHAMP spacecraft. However, in that study, Cluster spacecrafts were
in the exterior cusp and the distance between them and CHAMP were
about 60000km, while the distance between Cluster and CHAMP in our
case study is less than 20000km. \citet{B2012} also observed
that ions flow up from the ionosphere to the magnetosphere, apparently
as a consequence of field-aligned electron current which was coming
from the magnetosphere.
For both case studies in this paper, we have checked
densities and directions of oxygen ions measured from the CODIF CIS
instrument on Cluster, because these ions are usually coming from the
ionosphere.  No oxygen ions in periods of the formation of electron
current sheets could be detected. Most of the ions detected on the CIS
instrument were due to hydrogen which was originally coming from the
magnetosphere.
So at least in these two cases no ionospheric outflow was
detected. Without additional acceleration the ionospheric upflow would
ballistically return to the ionosphere \citep{O2009}, or it would take
the ions of the order of hours to reach an altitude of roughly
20000~km. Therefore we think that the failure to detect oxygen outflow
here is explained by the absence of energization of ions in the
topside ionosphere. This could be related to the low magnetic activity
conditions in these events. In addition, we have noticed that in one
event the Tsyganenko model field agreed well with the observed one,
while in the other there was a relatively large discrepancy between
model and observation in spite of low activity and low Cluster altitude.

\begin{acknowledgments}
  Tatjana \v{Z}ivkovi\'{c} would like to acknowledge the European
  Union Framework 7 Programme and the ECLAT project. Also, we
  acknowledge the ESA Cluster Archive and thank Eelco Doornbos for
  making a great web site from which we have downloaded CHAMP neutral
  density data. The DMSP particle detectors were designed by Dave Hardy and data
  obtained from JHU/APL.
\end{acknowledgments}

\end{article}
\end{document}